\documentstyle[12pt]{article}
\newcommand{\be}{\begin{equation}}
\newcommand{\ee}{\end{equation}}

\begin{document}
\baselineskip=14pt
\title{Stochastic models of exotic transport}
\author{Piotr Garbaczewski\thanks{Supported by KBN research
grant No 2 P03B 086 16}\\
Institute of Physics, Pedagogical University,  \\
  pl. S{\l}owia\'{n}ski 6, PL-65 069 Zielona G\'{o}ra,
  Poland}
\maketitle

\begin{abstract}
Non-typical  transport phenomena may arise when randomly driven
particles remain in an active relationship with the environment
instead of being passive. If we attribute to
Brownian particles an ability to induce  alterations of the environment
 on suitable space-time scales, those in turn must influence their
 further movement.   In that case a general  feedback
mechanism needs to be respected.
 By resorting to a specific choice of the
 particle-bath  coupling,  an enhanced (super-diffusion) or
 non-dispersive diffusion-type processes are found to exist in generically
 non-equilibrium contexts.
\end{abstract}

\section{Prerequisites}

A simple  Brownian motion, in its canonical (model)
manifestations, does not seem to hide any surprises. It is our
main goal in the present paper  to reconsider those ingredients of
the standard formalism which, while relaxed or slightly modified,
would lead to conceptually new and possibly exotic (in the sense
of being non-typical) features.

One of possible "defects" of the standard theory is that the
Brownian motion is  incapable to originate spatiotemporal patterns
(structures) and rather washes them out, even if initially in
existence. In this particular context, "active" Brownian particles
were introduced, \cite{geier}-\cite{schweitzer}. Most generally,
they are supposed to remain in a \it feedback \rm relationship
with the environment (heat reservoir, thermostat, random medium,
deterministic driving system or whatever else) through which they
propagate. If regarded as a thermal bath, the environment is
basically out of equilibrium.
Obviously, a concrete meaning of  \it  being active \rm  relies on
a specific choice of the model for the thermostat (deterministic
or random, linear or nonlinear etc.) and the detailed
particle-medium coupling mechanism.

Another "defective" feature of the standard theory is rooted in
its possible molecular (chaos) foundations where  the major
simplification is needed while  passing from the Boltzmann
collision theory to the novel kinetic framework set by the Kramers
equation, cf. \cite{dorfman}. In the process, microscopic
(molecular)  energy-momentum conservation laws completely
evaporate from the formalism. Then, there is no obvious way how to
reconcile  random  \it exchanges \rm of energy and momentum
between the Brownian particle and the thermostat with the manifest
isothermality requirement. An issue of the heat production/removal
and resulting thermal inhomogeneities  is normally ignored (cf.
however \cite{gar99}-\cite{gal1} for a discussion of various
thermostat models, and the notion of  \it thermostatting \rm  in
non-equilibrium dynamics).

That was the starting point of our discussion of the origins of
isothermal flows and the ultimate usage of the third Newton law in
the mean to justify the concept of  the Brownian motion \it with a
recoil \rm (via feedback relationship with the bath),
\cite{gar99}.

  In conformity with the metaphor, \cite{stanley}: "everything
depends on everything else", we relax the habitual sharp
distinction between an entirely passive particle (nonetheless
performing a continual erratic motion)and its exclusively active
(perpetually at rest and in equlibrium) driving agency - the
thermostat.

In the present paper we take the following point of view on major
signatures of the particle \it activity \rm while moving at random
due to external (environmental) reasons: any \it action \rm upon
the particle exerted by the environment induces (on suitable space
and time scales) a compensating \it reaction \rm in the carrier
random medium. In loose terms of Refs. \cite{geier}-\cite{lam} we
attribute to any (even single) Brownian particle  an ability to
generate perturbations of the medium (named a selfconsistent field
in Ref. \cite{geier} and perturbations of noise in Ref.
\cite{gar99}) which in turn influence its further movement. That
is a non-liner feedback relationship mentioned before: while
inducing a random dynamics of a particle (action), the medium
suffers alterations (reaction), which modify next stages of the
particle motion (feedback). Such features are  obviously alien to
the standard formalism.

There is no experimentally reliable way to watch Brownian
particles individually on time scales below say ${1\over {100}}s$,
nor get any insight into the associated fine detailed
particle-bath interaction phenomena. The only realistic procedure
to quantify observable features of the Brownian motion is to
exploit its familiar Janus face (cf. \cite{geier}). Namely, the
same mathematical formalism essentially applies to a single
particle  and to a statistical ensemble of identical
non-interacting Brownian particles. Then a  hydrodynamical analogy
can be exploited on both levels of description, \cite{klim,gar92},
once we invoke a standard kinetic reasoning (used in passing from
the Boltzmann equation to gas
 or liquid dynamics equations).

Effectively, we need to evaluate  (conditional) averages over a
statistical ensemble of particles, which (in loose terminology
that follows Ref. \cite{klim}, see also \cite{ebeling})
constitutes one component (noninteracting Brownian "gas") of a
coupled two-component continuous system. Another component is the
thermostat.

What are  the ultimate kinetic features of  the Brownian motion
(diffusion process) critically relies on the thermostat model i.e.
specific  mechanisms of the energy/heat exchange due to the
coupling between the thermostat and the Brownian "gas".
 This particular issue we shall investigate in some detail in
below, by resorting to somewhat nontypical (exotic in the Brownian
motion context) methods.

\section{Local conservation  laws for the Brownian motion
and  Smoluchowski-type diffusion}

It is useful  to exploit a standard phase-space argument that is
valid, under isothermal conditions,  for a Markovian diffusion
process taking place in (or relative to) a driving  flow
$\vec{w}(\vec{x},t)$ with as yet unspecified dynamics nor concrete
physical origin. (In particular, such a flow can receive an
interpretation of a selfconsistent field generated in the
environment by Brownian particles themselves, cf. Ref.
\cite{geier}.)

We account for an explicit external force (here, acceleration
$\vec{K}= \vec{F}/m$) exerted upon diffusing particles, while not
directly affecting the driving flow itself. Then,  infinitesimal
increments of phase-space random variables read: \be d\vec{X}(t)=
\vec{V}(t) dt \ee \be {d\vec{V}(t)= \beta [\vec{w}(\vec{x},t) -
\vec{V}(t)] dt + \vec{K}(\vec{x})dt  + \beta \sqrt{2D}
d\vec{W}(t)\enspace .} \ee

Following the leading idea of the Smoluchowski approximation, we
assume
 that $\beta $ is large, and consider the process on  time scales
 significantly exceeding $\beta ^{-1}$.
 Then, an appropriate choice of
 the velocity field $\vec{w}(\vec{x},t)$ may
 in principle guarantee  the convergence of the
 spatial part
 $\vec{X}(t)$ of the process  to  the It\^{o} diffusion process
 with infinitesimal increments :
 \be {d\vec{X}(t) = \vec{b}(\vec{x},t) dt
 +  \sqrt{2D} d\vec{W}(t)\enspace .}\ee
 In this case,  the forward drift  of the process  reads
$\vec{b}(\vec{x},t)= \vec{w}(\vec{x},t) + {1\over \beta }
\vec{K}(\vec{x})$. Notice that the
 $\beta ^{-1}\vec{K}$ contribution  can be safely ignored if
 we are interested in the  dominant driving motion.

Throughout the paper we are  interested in  Markovian
 diffusion processes, which propagate respectively the
 phase-space or
 configuration space probability densities (weak solutions
 of stochastic differential equations are thus involved).
 In the configuration space
 variant, we deal with  a  Markovian  stochastic  process
 whose probability
density $\rho (\vec{x},t)$ evolves according to  the standard
Fokker-Planck equation
 \be {\partial _t \rho =
 D\triangle  \rho - \vec{\nabla }\cdot (\vec{b}\rho )}
\ee and the forward drift  is not necessarily of the form dictated
by Eqs. (1) and  (2). We admit here more general forward drift
functions, cf. Ref. \cite{gar99}, which do not  allow for a simple
additive decomposition  and thus are capable to give account of
nonlinearities in the particle-bath coupling.

One can easily transform the Fokker-Planck equation to the
familiar form of the continuity equation (hydrodynamic mass
conservation law) ${\partial _t\rho =
- \vec{\nabla }\cdot (\vec{v} \rho )}$  by defining
 $\vec{v}=\vec{b} - D{{\vec{\nabla }\rho }
 \over {\rho }}$.
The current velocity $\vec{v}$ obeys a local (momentum per unit of
mass) conservation law  which  directly originates from the rules
of the It\^{o} calculus for Markovian diffusion processes,
 and from the first moment equation in the
diffusion approximation  of the Kramers theory, \cite{gar99,gar92}:
\be {\partial _t\vec{v} + (\vec{v} \cdot \vec{\nabla }) \vec{v} =
\vec{\nabla }(\Omega - Q)\enspace .}\ee

While looking similar to the standard Euler equation appropriate
for the lowest order hydrodynamical  description of gases and
liquids, this equation  conveys an  entirely different physical
message.

First of all,  for a class of forward  drifts that
are gradient fields   the most  general admissible
 form of an auxiliary potential $\Omega (\vec{x},t)$ reads:
\be {\Omega (\vec{x},t) = 2D[ \partial _t\phi + {1\over 2}
({\vec{b}^2\over {2D}} + \vec{\nabla }\cdot \vec{b})]\enspace .}
\ee Here $\vec{b}(\vec{x},t) = 2D \vec{\nabla } \phi (\vec{x},t)$.
In reverse, by choosing a bounded from below continuous function
to represent conservative  force fields (after taking the
gradient) i. e.  otherwise arbitrary  $\Omega $, we can always
disentangle  the above (Riccatti-type) identity with respect to
the drift field.

 Moreover, instead of the standard pressure term (consider a
 state  equation  $P\sim \rho ^\alpha , \alpha >0$),
 there appears a contribution from more complicated (derivatives !)
    $\rho $-dependent potential
 $Q(\vec{x},t)$. It is  is given in terms of the so-called osmotic
velocity field  $\vec{u}(\vec{x},t) = D\vec{\nabla } \,
ln \rho (\vec{x},t)$:
\be {Q(\vec{x},t) = {1\over
2} \vec{u}^2 + D\vec{\nabla } \cdot \vec{u}} \enspace .\ee

 An equivalent form of the enthalpy-related
potential $Q$ is
 $Q= 2D^2{{\triangle \rho ^{1/2}} \over {\rho ^{1/2}}}$.

A  general expression for the local diffusion current is $\vec{j}
= \rho \vec{v} = \rho (\vec {b} - D{{\vec{\nabla }\rho }\over
{\rho }})$. This  local flow in principle may  be experimentally
observed for  a cloud of suspended particles in a liquid.  The
current  $\vec{j}$ is nonzero  in  non-equilibrium situations and
a non-negligible matter transport occurs as a consequence of the
Brownian motion, on the ensemble average. We thus  cannot avoid
local heating/cooling phenomena that need to push the environment
out of equilibrium. That leads to obvious temperature
inhomogeneities, which are normally disregarded, cf. Ref.
\cite{gar99}.

If the forward  drift is  interpreted as  a gradient of a suitable
function ($b=2D\nabla \phi =\nabla \Phi $)  and we take $\Phi
(\vec{x},0)$ as the initial data for the $t\geq 0$ evolution),
then we have:
 \be {\Omega =
\partial _t \Phi + {1\over 2}|\vec{\nabla } \Phi |^2 + D\triangle
\Phi \enspace .}\ee

If we decide that  the above Hamilton-Jacobi-type equation
 is to be solved with respect to the field
$\Phi (\vec{x},t)$, its  solution (and general solvability issue)
relies on the choice of a bounded from below, continuous function
$\Omega (\vec{x},t)$, \it without \rm any a priori knowledge of
forward drifts. Viewed that way,
 Eq. (7) sets limitations on admissible forms of the space-time
 dependence of
any conceivable self-consistent  field to be generated in the bath
by Brownian particles  (cf. Ref. \cite{geier}). There is no
freedom at all for postulating various partial differential
equations, if their solutions are to be interpreted as forward
drifts of Markovian diffusion processes, \cite{gar98}.

There  is also  interesting to observe that a gradient
field ansatz for the diffusion current velocity
$\vec{v}=\vec{\nabla }S$  allows
to transform the momentum conservation law  of a Markovian
diffusion process to the universal Hamilton-Jacobi form: \be
{\Omega =
\partial _tS + {1\over 2} |\vec{\nabla }S|^2  + Q } \ee
where $Q(\vec{x},t)$ was defined before as the
"pressure/enthalpy"-type  function. (That form looks deceivingly
similar to the  standard hydrodynamic conservation law, valid for
liquids and gases at thermal equilibrium,  where instead of $Q$ an
enthaply function normally appears.)
 By performing  the gradient
operation  we recover the previous hydrodynamical form of the law.

In the above, the contribution due to $Q$ is a direct consequence
of an initial probability measure choice for the diffusion
process,  while $\Omega $  alone does account for an appropriate
forward drift of the process, playing at the same time the role of
the volume force potential ($\nabla Q ={\nabla P\over \rho }$
contributes to   energy-momentum transfer effects through the
boundaries of any volume).

\section{Moment equations for the free Brownian motion:
Getting out of conventions}

 The  derivation of a hierarchy of local conservation laws
(moment equations)  for the Kramers equation can be patterned
after the standard procedure for the Boltzmann equation,
\cite{dorfman,gar92,klim}. Those laws do not form a closed system
and additional specifications (like the familiar thermodynamic
equation of state) are needed to that end. In case of the
isothermal Brownian motion, when considered in the large friction
regime (e.g. Smoluchowski diffusion approximation), it suffices to
supplement  the Fokker-Planck equation  by one more conservation
law \it only \rm to arrive at  a closed system.

To give a deeper insight into what really happens on the way from
the phase-space theory of the Brownian motion to its approximate
configuration-space (Smoluchowski) version, let us consider the
familiar Ornstein-Uhlenbeck process (in velocity/momentum)
in its extended phase-space form.     For clarity of discussion,
we discus random dynamics for one degree of freedom only.

In the absence of external forces, the kinetic
(Kramers-Fokker-Planck equation) reads: \be {\partial _t W +
u\nabla _xW = \beta \nabla _u(Wu) + q \triangle _uW}  \ee where
$q=D\beta ^2$.   Here  $\beta $  is the friction coefficient, $D$
will be identified later with the  spatial diffusion constant, and
provisionally we set $D=k_BT/m\beta $ in conformity with  the
Einstein fluctuation-dissipation identity.

The  joint probability distribution (in fact, density) $W(x,u,t)$
for a freely moving Brownian particle which at $t=0$ initiates its
motion at $x_0$ with an arbitrary inital velocity $u_0$ can  be
given in the form of the maximally symmetric displacement
probability law:
\be {W(x,u,t)=W(R,S) = [4\pi ^2(FG-H^2)]^{-1/2}
\cdot } exp\{ - {{GR^2 - HRS + FS^2}\over {2(FG - H^2)}}\}\ee
where $R=x-u_0(1-e^{-\beta t})\beta ^{-1}$, $S=u-u_0e^{-\beta t}$
while $ F = {D\over \beta }(2\beta t - 3 +4e^{-\beta t}-
e^{-2\beta t})$\, $G=D\beta (1-e^{-2\beta t})$ and
$H=D(1-e^{-\beta t})^2$.

Marginal  probablity densities, in the Smoluchowski regime
(we take for granted  that time scales $\beta ^{-1}$ and space scales
$(D\beta ^{-1})^{1/2}$ are irrelevant),  take familiar
forms of the  Maxwell-Boltzmann
$w(u,t)=({m\over {2\pi kT}})^{1/2}\, exp( - {{mu^2}\over {2k_BT}})$
and the diffusion kernel
$w(x,t) =(4\pi Dt)^{-1/2} exp( - {x^2\over {4Dt}})$ respectively.

A direct evaluation of the first and second local moment
of the phase-space probability density:
\be <u> = \int du\, uW(x,u,t)= w(R) [(H/F)R + u_0e^{-\beta t}] \ee
\be { <u^2> = \int du \,  u^2W(x,u,t) =
({{FG-H^2}\over F}+ {H^2\over F^2}R^2)
\cdot }(2\pi F)^{-1/2} exp(- {R^2\over {2F}})\ee
after  passing to the diffusion (Smoluchowski) regime, \cite{gar92},
 allows  to  recover
the  local (configuration space conditioned) moment
 $<u>_x={1\over w}<u>$ which reads
\be <u>_x = {x\over {2t}}= - D{{\nabla w(x,t)}\over {w(x,t)}} \ee
while for  the second local moment $<u^2>_x= {1\over w}<u^2>$
 we arrive at
\be {<u^2>_x = (D\beta - D/2t) + <u>^2_x \enspace .}\ee

By inspection one verifies that the transport (Kramers)
 equation  for $W(x,u,t)$ implies local conservation laws:
 \be \partial _t w + \nabla (<u>_x w) = 0 \ee
 and
 \be {\partial _t(<u>_x w) + \nabla _x(<u^2>_xw) = - \beta <u>_x w
 \enspace .}\ee

By introducing (we strictly follow the moment equations strategy
of the traditional kinetic theory of gases and liquids)  the
notion of pressure function  $P_{kin}$ (we choose another notation
to make a difference with the previous notion $P$  of the pressure
function, cf. $\nabla Q= {{\nabla P}\over \rho }$): \be
P_{kin}(x,t) = (<u^2>_x - <u>^2_x) w(x,t) \ee we can analyze the
local momentum conservation law \be (\partial _t + <u>_x \nabla
)<u>_x = - \beta <u>_x - {{\nabla P_{kin}} \over w} \ee

In the Smoluchowski regime  the friction term is cancelled away by
a counterterm  coming from ${1\over w} \nabla P_{kin}$  so that
\be (\partial _t + <u>_x\nabla )<u>_x = {D\over {2t}}{{\nabla w}\over w} =
- {{\nabla P}\over w} = - \nabla Q \ee
where $P=D^2w\triangle ln\, w$.

For comparison with the notations (local conservation laws)
of the previous section one needs
to replace $w(x,t)$ by $\rho (x,t)$ and $<u>_x$ by $v(x,t)$ in all
 formulas that pertain to the Smoluchowski regime.

Further exploiting the kinetic lore, we can   tell  few words
about the \it  temperature of Brownian particles \rm as opposed to
the (posibly equilibrium) temperature of the thermal bath. Namely,
in view  of  (we stay in the Smoluchowski regime) $P_{kin} \sim
(D\beta - {D\over {2t}})w$ where $D={{k_BT}\over m{m\beta }}$, we
can formally  set: \be { \Theta = {P_{kin} \over w} \sim (k_BT -
{D\over {2t}}) < k_BT }\enspace .\ee

That quantifies the degree of thermal agitation (temperature) of
Brownian particles to be \it less \rm than the thermostat
temperature. Heat is continually pumped from the  thermostat to
the Brownian "gas", until asymptotically both temperatures
equalize. This  may  be called a "thermalization" of Brownian
particles. In the process of thermalization the   Brownian "gas"
temperature monotonically  grows up until  the mean kinetic energy
of particles     and that of mean flows asymptotically approach
the familiar kinetic relationship: \be \int  {w\over 2}(<u^2>_x -
<u>^2_x) dx = k_BT \ee

In view of this medium $\rightarrow $ particles  heat transfer
issue, one must be really careful while associating habitual
thermal equilibrium conditions with essentially non-equilibrium
phenomena, cf. Ref. \cite{gar99} for more extended discussion.

\section{Hydrodynamical reasoning: In search for exotic}

Once local conservation laws were introduced, it seems instructive
to comment on the essentially hydrodynamical features
(compressible fluid/gas case) of the problem. Specifically,  the
"pressure" term $\nabla Q$ is here quite annoying from the
traditional kinetic theory perspective. That is quite apart from
the fact that our local conservation laws have a conspicuous Euler
form appropriate for the standard hydrodynamics of gases and
liquids. Let us stress that in case of normal liquids the pressure
is exerted upon any control volume (droplet) by the surrounding
fluid. We may interpret that as a  compression of a  droplet. In
case of Brownian motion, we deal with a definite decompression:
particles are driven away from areas of higher concentration
(probability of occurence). Hence, typically the  Brownian
"pressure" is exerted by the droplet upon its surrounding.

Following the hydrodynamic tradition let us analyze that
"pressure" issue in more detail. We consider a reference volume
(control interval, finite droplet) $[\alpha ,\beta ]$ in $R^1$ (or
$\Lambda \subset R^1$ ) which at time $t\in [0,T]$ comprises a
certain fraction of particles (Brownian "fluid" constituents). The
time rate  of particles loss  or gain by the volume $[\alpha,\beta
]$ at time $t$, is equal to the flow outgoing through the
boundaries i.e. \be -\partial _t \int_{\alpha }^{\beta }\rho
(x,t)dx = \rho (\beta ,t)v(\beta ,t) - \rho (\alpha ,t)v(\alpha
,t) \ee which is a consequence of the continuity equation.

To analyze the momentum balance, let us allow for an infinitesimal
deformation of the boundaries of $[\alpha ,\beta ]$ to have
entirely compensated the mass (particle) loss or gain: \be [\alpha
,\beta ] \rightarrow [\alpha +v(\alpha ,t)\triangle t,\beta
+v(\beta ,t)\triangle t]  \ee

 Effectively, we pass then to the locally co-moving frame.
That implies
\be lim_{\triangle t\downarrow 0} {1\over {\triangle
t}}\bigl [\int _{\alpha +v_{\alpha }\triangle t}^{\beta
+v_{\beta}\triangle t} \rho (x,t+\triangle t)dx - \int_{\alpha
}^{\beta } \rho (x,t)dx\bigr ] =
 \ee
$$= lim_{\triangle t \downarrow 0}{1\over {\triangle t}}\bigl [\int_{\alpha +
v_{\alpha
}\triangle t}^{\alpha } \rho (x,t)dx + \triangle t\int_{\alpha }^{\beta
}(\partial _t\rho ) dx + \int_{\beta }^{\beta +v_{\beta }\triangle t}
\rho (x,t) dx\bigr ]=0$$

Let us investigate what happens to the local matter  flows
$(\rho v)(x,t)$, if we proceed in the same way (leading terms only
are retained): \be \int_{\alpha +v_{\alpha }\triangle t}^{\beta
+v_{\beta }\triangle t} (\rho v)(x,t+\triangle t)dx - \int_{\alpha
}^{\beta }(\rho v)(x,t) dt \sim   \ee
 $$\sim -(\rho v^2)(\alpha
,t)\triangle t + (\rho v^2)(\beta ,t)\triangle t +\triangle t
\int_{\alpha }^{\beta }[\partial _t (\rho v)]dx$$

In view of local conservation laws  we have $\partial _t(\rho
v)=-\nabla (\rho v^2) +\rho \nabla (\Omega -Q)  $ and the rate of
change of momentum associated with the control volume $[\alpha
,\beta ]$ is  (here per unit of mass)
\be lim_{\triangle
t\downarrow 0}{1\over {\triangle t}}\bigl [\int_{\alpha +
v_{\alpha }\triangle t}^{\beta +v_{\beta }\triangle t} (\rho
v)(x,t+\triangle t)- \int_{\alpha }^{\beta } (\rho v)(x,t)\bigr ]=
\int_{\alpha }^{\beta } \rho \nabla (\Omega -Q) dx    \ee

However, $\nabla Q = {{\nabla P}\over \rho }$ and $P=D^2\rho
\triangle ln \rho $. Therefore:
\be \int_{\alpha }^{\beta }\rho \,
\nabla (\Omega -Q)dx = \int_{\alpha }^{\beta } \rho \nabla \Omega
dx - \int_{\alpha }^{\beta } \nabla P dx =   \ee
 $$=E[\nabla
\Omega ]_{\alpha }^{\beta } + P(\alpha ,t) - P(\beta ,t) $$

Clearly, $\nabla \Omega $ refers to the Euler-type volume force,
while $\nabla Q$ (or more correctly, $P$) refers to the "pressure"
effects entirely due to the particle transfer rate through the
boundaries of the considered volume.

The missing  ingredient of our discussion is the time developement of
 the kinetic
energy of the matter flow ${1\over 2} (\rho v^2)$  (per unit of
mass) transported through the chosen volume. Let us therefore
evaluate: \be \int_{\alpha +v_{\alpha }\triangle t}^{\beta
+v_{\beta }\triangle t} {1\over 2}(\rho v^2)(x,t+\triangle t)dx -
\int_{\alpha }^{\beta }{1\over 2} (\rho v^2)(x,t) dt \sim   \ee
 $$\sim -{1\over 2}(\rho v^3)(\alpha
,t)\triangle t + {1\over 2}(\rho v^3)(\beta ,t)\triangle t +\triangle t
\int_{\alpha }^{\beta }[\partial _t {1\over 2}(\rho v^2)]dx$$

We have  $ \partial _t ({1\over 2}\rho v^2) = - {1\over 2}
v^2\nabla (\rho v) - \rho v \nabla (Q-\Omega + {1\over 2}v^2) =
 - \nabla ({1\over 2}\rho v^3) -
 \rho v\nabla (Q-\Omega)$.
Consequently,  the time rate of the  kinetic energy (of the flow)
loss/gain by the volume reads: \be lim_{\triangle t\downarrow
0}{1\over {\triangle t}}\bigl [\int_{\alpha + v_{\alpha }\triangle
t}^{\beta +v_{\beta }\triangle t} {1\over 2} (\rho
v^2)(x,t+\triangle t)- \int_{\alpha }^{\beta } {1\over 2}(\rho
v^2) (x,t)\bigr ]= \int_{\alpha }^{\beta } (\rho v) \nabla (\Omega
-Q) dx    \ee

In the integrand one immediately recognizes an expression (up to the
mass parameter)    for the power released/absorbed by the control
volume at time $t$.

In particular, by taking advantage of the identity $\nabla Q=
{{\nabla P}\over \rho }$ we can rewrite the pure "pressure"
 contribution as follows:  $-\int_{\alpha }^{\beta }  \rho v\nabla Qdx =
 - \int_{\alpha }^{\beta }v\nabla Pdx$ which clearly is a direct
  analog of the standard mechanical expression for the power
 release (${{dE}\over {dt}} = F\cdot v$).

Surely, one cannot interpret the above local averages as an
outcome of an innocent operation of the random medium upon
particles. There is  a non-negligible transfer of energy and
momentum to be accounted for  in association with  the Brownian
motion, and a  number of problems  (when can we disregard
temperature gradients ?) pertaining to local heating and cooling
phenomena suffered by the environment (in terms of local averages)
should be consistently   resolved.

\section{Implementing the feedback}

In contrast to molecular chaos derivations based on the Boltzmann
equation, \cite{dorfman}, in case of the Brownian motion we have
no access into the microscopic fine details of particle
propagation. Instead, we need to postulate the mathematically
reliable form of the dynamics, even if ending up with obvious
artifacts of the formalism (like  e.g. non-differentiable paths,
or their infinite spatial variation on arbitrarily short time
intervals, in case of the Wiener process or, even worse,
white-noise input).

Our major hypothesis about an exotic particle-bath coupling
pertains to local averages and is motivated by a fairly intuitive
picture of the behaviour of suspended particles in thermally
inhomogeneous media. Namely, it is well known that hot areas
support much lower density of suspended particles (can even be
free of any dust admixtures) than the lower temperature areas.
Dynamically we can interpret this phenomenon as a \it repulsion
\rm of suspended particles by warm areas and an \it  attraction
\rm by the cool ones, cf. Ref. \cite{gar99}.

In the course of our  discussion we have spent quite a while on
demonstrating that a non-trivial energy/heat exchange is
completely ignored in the standard approach to the Brownian motion
(even, if under suitable circumstances one has good reasons to do
that). Effectively, the model dramatically violates basic
conservation laws of physics. That derives from the assumption
that the thermostat is perpetually in the state of  rest (in the
mean), in thermal equilibrium  and thus  free of any thermal
currents.

The force/acceleration term  (we turn back to the
three-dimensional notation)  $\vec{\nabla }(\Omega - Q)$ appears
in all formulas that refer to  energy/momentum flows supported by
local averages of the  Brownian motion. Their sole reason is the
\it action \rm of the random medium upon particles which causes an
expansion of the Brownian "swarm" out of  areas of higher
concentration. That however needs a local cooling of the medium
(Brownian particles are being "thermalized" by increasing their
mobility e.g. temperature). On sufficiently low time scales that
should amount to an instantaneous  \it reaction \rm of the medium
in terms of thermally induced currents: Brownian particles are
driven back (attraction !) to the cooler areas i.e. float \it down
\rm the temperature gradients as long as they are non-vanishing,
 \cite{widder}.

We shall bypass the thermal inhomogeneity issue by resorting to an
explicit energy-momentum balance information available through
local conservation  laws for the Brownian motion.

 If we regard the term $\vec{\nabla }(\Omega - Q)$ as a quantification
of the sole medium \it action  \rm upon particles, then the most
likely quantification of the medium \it reaction \rm should be
exactly the opposite i.e. $\vec{\nabla }(Q-\Omega)$.

Told otherwise, medium "effort" to release momentum and energy
from the Brownian "droplet" (control volume),  at a time rate
determined by the functional form of $\vec{\nabla }(\Omega
-Q)(x,t)$, induces a compensating (in view of the heat deficit)
energy and momentum delivery to that volume needed to remove the
thermal inhomogeneity of the thermostat. As a consequence,
Brownian particles propagate through  the medium which is non
longer in the state of rest and develops intrinsic (mean, on the
ensemble average) flows.

A mathematical encoding of this Brownian recoil principle
\cite{vigier}, or third Newton law in the mean \cite{gar99}
hypothesis is rather well established.

The momentum conservation law for the process \it with a recoil
\rm (the reaction term replaces the decompressive  action  term)
will read: \be {\partial _t\vec{v} + (\vec{v}\cdot \vec{\nabla
})\vec{v} = \vec{\nabla } (Q- \Omega )}\ee
 implying that
\be {\partial _t S + {1\over 2} |\vec{\nabla }S|^2 - Q= -\Omega
}\ee
 stands for the corresponding Hamilton-Jacobi equation, cf.
\cite{zambrini}, instead of "normal" one. A suitable adjustment
(re-setting) of the initial data is here necessary, cf.
\cite{gar99}.

In the coarse-grained picture of motion we shall deal with a
sequence    of repeatable feedback scenarios realized on the
Smoluchowski process  time scale: the Brownian "swarm" expansion
build-up is accompanied by the parallel counter-flow build-up,
which in turn modifies the subsequent stage of the Brownian
"swarm" migration (being interpreted to modify the forward drift
of the process) and the corresponding, built-up anew counter-flow.

Perhaps surprisingly, we are still dealing with Markovian
diffusion-type processes, \cite{nel}. The link is particularly
obvious \cite{gar96} if we observe that the
 new Hamilton-Jacobi equation  can be formally
rewritten in the previous form  by introducing:
\be {\Omega _r=
\partial _tS + {1\over 2} |\vec{\nabla }S|^2 + Q }\ee
where ${\Omega _r= 2Q - \Omega }$ and $\Omega $ represents the
previously defined potential function characterizing any
Smoluchowski (or more general) diffusion process. It  is however
$\Omega _r$  which would determine forward drifts of the Markovian
diffusion process with a recoil. Those  must come out from  the
Cameron-Martin-Girsanov identity $ \Omega _r = 2Q- \Omega = 2D[
\partial _t\phi + {1\over 2} ({\vec{b}^2\over {2D}} + \vec{\nabla
}\cdot \vec{b})]$.

After complementing the Hamilton-Jacobi-type equation by the
continuity equation, we again end up with a closed system of
conservation laws. The  system is badly nonlinear and coupled, but
its linearisation can be immediately given in terms of an adjoint
pair of Schr\"{o}dinger equations with  a potential $\Omega $ (the
imaginary unit $i$ on the left-hand-side in below is not an error
!), \cite{nel,zambrini,gar96}. Indeed, \be {i\partial _t \psi = -
D\triangle \psi + {\Omega \over {2D}}\psi } \ee
 with a solution represented in the polar form ${\psi =
\rho ^{1/2} exp(iS)}$ and its complex adjoint makes the job.
 The
choice of $\psi (\vec{x},0)$ sets here  a solvable Cauchy problem.
 Notice that, in view of the Schr\"{o}dinger-type linearization,
  for time-indepedent $\Omega $ (conservative forces)
the total energy $\int_{R^3}({v^2\over 2} -Q + \Omega )\rho d^3x$
 of the system  is a  conserved finite quantity.
We thus reside within the framework of so-called \it finite energy
\rm diffusion processes, whose mathematical features  received
some attention in the literature. In particular, it is known that
in the absence of volume forces, a superdiffusion appears, while
harmonic volume forces  allow for non-dispersive diffusion-type
processes, \cite{gar99}.

\end{document}